\documentclass[journal]{IEEEtran}

\usepackage{color}
\usepackage{balance}
\usepackage{graphicx}
\usepackage{enumerate}
\usepackage{subfigure}
\usepackage{amsmath}
\usepackage{algorithm} 
\usepackage{algorithmic}
\usepackage{booktabs}
\usepackage{cite}
\usepackage{bm}
\usepackage{amssymb}

\begin{document}

\title{DeepGOMIMO: Deep Learning-Aided Generalized Optical MIMO with CSI-Free Blind Detection}

\author{
Xin Zhong,
Chen Chen, \IEEEmembership{Member, IEEE,}
Shu Fu,
Zhihong Zeng,
and Min Liu

\thanks{This work was supported in part by the National Natural Science Foundation of China under Grant 61901065, in part by the Fundamental Research Funds for the Central Universities under Grant 2021CDJQY-013 and Grant 2020CDJQY-A001, and in part by the China Postdoctoral Science Foundation under Grant 2021M693744.} 

\thanks{X. Zhong, C. Chen, S. Fu, and M. Liu are with the School of Microelectronics and Communication Engineering, Chongqing University, Chongqing 400044, China (e-mail: \{c.chen, 201912131098, shufu, liumin\}@cqu.edu.cn).}

\thanks{Z. Zeng is with the LiFi Research and Development Centre, Institute for Digital Communications, The University of Edinburgh, EH9 3JL, UK (e-mail: zhihong.zeng@ed.ac.uk).}
}



\maketitle

\begin{abstract}
Generalized optical multiple-input multiple-output (GOMIMO) techniques have been recently shown to be promising for high-speed optical wireless communication (OWC) systems. In this paper, we propose a novel deep learning-aided GOMIMO (DeepGOMIMO) framework for GOMIMO systems, where channel state information (CSI)-free blind detection can be enabled by employing a specially designed deep neural network (DNN)-based MIMO detector. The CSI-free blind DNN detector mainly consists of two modules: one is the pre-processing module which is designed to address both the path loss and channel crosstalk issues caused by MIMO transmission, and the other is the feed-forward DNN module which is used for joint detection of spatial and constellation information by learning the statistics of both the input signal and the additive noise. Our simulation results clearly verify that, in a typical indoor 4 $\times$ 4 MIMO-OWC system using both generalized optical spatial modulation (GOSM) and generalized optical spatial multiplexing (GOSMP) with unipolar non-zero 4-ary pulse amplitude modulation (4-PAM) modulation, the proposed CSI-free blind DNN detector achieves near the same bit error rate (BER) performance as the optimal joint maximum-likelihood (ML) detector, but with much reduced computational complexity. Moreover, since the CSI-free blind DNN detector does not require instantaneous channel estimation to obtain accurate CSI, it enjoys the unique advantages of improved achievable data rate and reduced communication time delay in comparison to the CSI-based zero-forcing DNN (ZF-DNN) detector.
\end{abstract}


\section{Introduction}
\IEEEPARstart{D}{ue} to the exhaustion of radio frequency (RF) spectrum resources, optical wireless communication (OWC) which explores the infrared, visible light or ultra-violet spectrum has been envisioned as a promising candidate to satisfy the ever-increasing data demand in future indoor environments \cite{ghassemlooy2015emerging}. In recent years, bidirectional OWC, which is also named light fidelity (LiFi), has been widely considered as one of the key enabling technologies for 5G/6G and Internet of Things (IoT) communications \cite{cogalan2017would,chi2020visible,demirkol2019powering,chen2021noma}. Although OWC systems have many inherent advantages such as abundant license-free spectrum resources, no electromagnetic interference (EMI) and enhanced physical-layer security, the practically achievable capacity of OWC systems is largely limited by the small modulation bandwidth of commercial off-the-shelf (COTS) optical elements, especially for illumination light-emitting diodes (LEDs) \cite{le2009100}.

As a very natural way to efficiently improve the achievable capacity of indoor OWC systems using LEDs, multiple-input multiple-output (MIMO) transmission has attracted great attention recently, which fully exploits the existing LED fixtures in the ceiling of a typical room to harvest substantial diversity or multiplexing gain \cite{zeng2009high,fath2013performance,chen2017coverage}. So far, various optical MIMO techniques have been introduced for OWC systems, among which optical spatial multiplexing (OSMP) and optical spatial modulation (OSM) are two most popular ones. Specifically, OSMP can achieve full multiplexing gain and hence a relative high spectral efficiency, but suffers from severe inter-channel interference (ICI) \cite{chen2020user}. In contrast, OSM can remove ICI by activating a single LED to transmit signal at each time slot. Although OSM can transmit additional index bits, only one constellation symbol can be transmitted at each time slot, and hence it is challenging for OSM systems to achieve high spectral efficiency \cite{mesleh2011optical}. Lately, generalized optical MIMO (GOMIMO) techniques, including generalized OSM (GOSM) and generalized OSMP (GOSMP), have been further proposed to boost the capacity of MIMO-OWC systems \cite{alaka2015generalized,wang2020constellation,wang2020indoor,chen2021ofdm}. In GOSM systems, multiple LEDs are activated to transmit the same signal, and therefore more index bits can be transmitted and the diversity gain can also be increased. In GOSMP systems, only a subset of LEDs are activated to transmit different signals, resulting in reduced multiplexing gain. However, additional index bits can be transmitted and the ICI can also been reduced in GOSMP systems. Our previous work \cite{chen2021ofdm} has already clearly demonstrated the superiority of GOMIMO techniques in comparison to conventional OSM and OSMP.

In order to successfully implement GOMIMO systems, an efficient MIMO detection scheme should be adopted. Generally, the joint maximum-likelihood (ML) detector serves as the optimal detector for GOMIMO systems \cite{ozbilgin2015optical}. Nevertheless, the ML detector usually has high computational complexity, making it infeasible in practical applications. Instead, the combination of zero-forcing (ZF) equalization and ML detection can be a practical low-complexity detection scheme for GOMIMO systems \cite{ozbilgin2015optical}. However, ZF equalization inevitably leads to noise amplification due to high channel correlation in typical indoor MIMO-OWC systems. Moreover, the ZF-ML detector also suffers from the adverse effect of error propagation, since the detection error of spatial symbols might propagate to the estimation of constellation symbols.

With the rapid development of machine learning technology, deep learning has revealed its great potential in wireless communication systems \cite{lecun2015deep,wang2017deep}. Most recently, deep learning techniques have also been applied in OWC systems for binary signaling design \cite{lee2018binary}, mitigation of both linear and nonlinear impairments \cite{lu2019memory}, energy-efficient resource management \cite{yang2019learning}, and so on. More specifically, a ZF-based deep neural network (DNN) detection scheme has been proposed for MIMO detection in GOMIMO systems \cite{wang2020deep}. The obtained results in \cite{wang2020deep} show that the ZF-DNN detector can achieve comparable  bit error rate (BER) performance as the optimal joint ML detector with greatly reduced computational complexity. Nevertheless, the ZF-DNN detector takes the ZF equalized signal as its input, which requires accurate channel state information (CSI), i.e., the MIMO channel matrix, to successfully perform ZF equalization. Although CSI can be estimated by using training symbols \cite{wang2014demo}, training-based instantaneous channel estimation inevitably causes both the loss of achievable data rate and the increase of communication time delay.

\begin{figure*}[!t]
\centering
{\includegraphics[width=1.6\columnwidth]{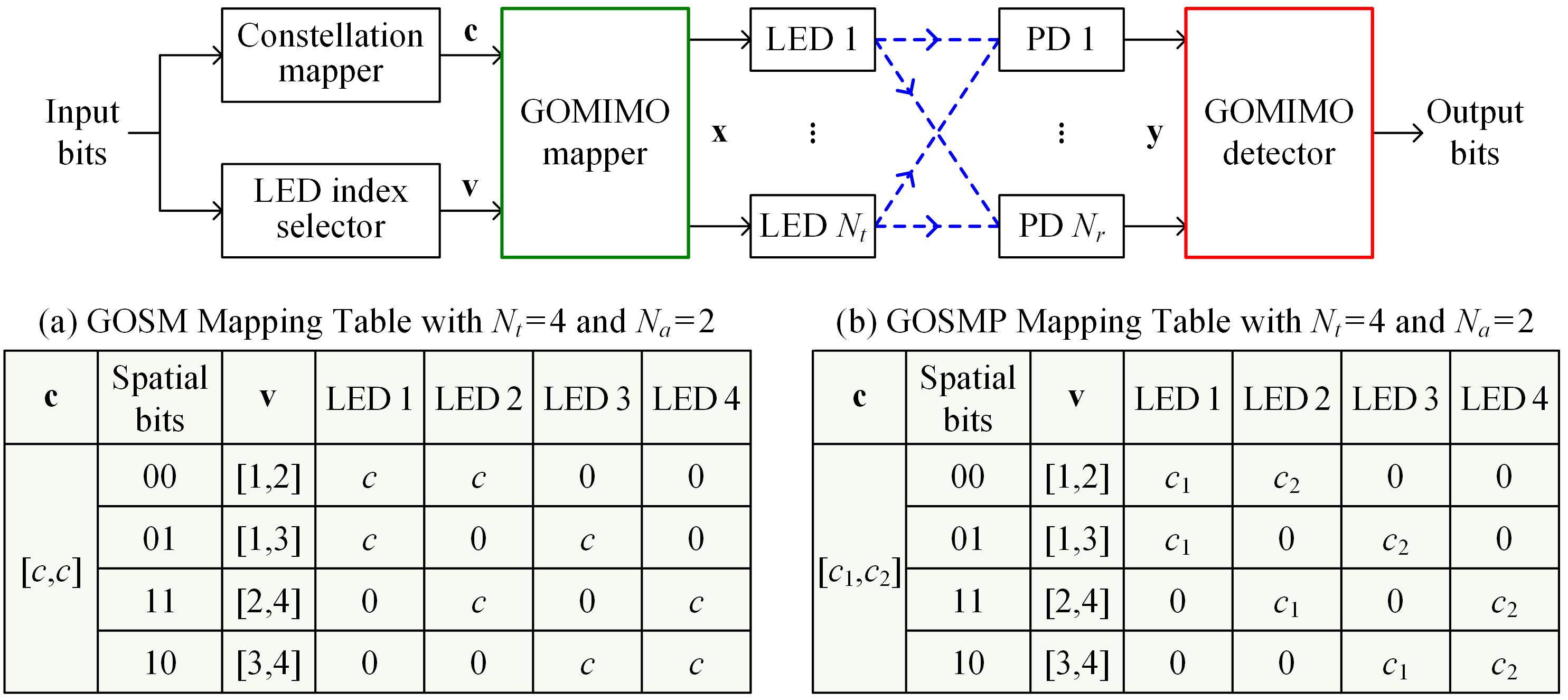}} 
\caption{Block diagram of a general $N_r \times N_t$ GOMIMO system. Insets (a) and (b) show the mapping tables of GOSM and GOSMP, respectively.}
\label{fig:diagram}
\end{figure*}

In this paper, to address the disadvantages of CSI-based ZF-DNN detection due to the requirement of instantaneous CSI for ZF equalization, we for the first time propose a DeepGOMIMO framework for GOMIMO systems where CSI-free MIMO detection is achieved by a novel blind DNN detection scheme. By adding a specially designed pre-processing module before the feed-forward DNN module, CSI-free blind detection can be successfully enabled for GOMIMO systems. Numerical simulations are extensively conducted to evaluate the performance of the proposed CSI-free blind DNN detector, which is also compared with other three benchmark schemes including the joint ML detector, the ZF-ML detector and the ZF-DNN detector. Our simulation results verify the advantages of the proposed CSI-free blind DNN detector in comparison to other benchmark schemes in GOMIMO systems.

The rest of this paper is organized as follows. In Section II, we describe the mathematical model of a general GOMIMO system. In Section III, we introduce four detection schemes for GOMIMO systems. Detailed simulation setup and results are presented in Section IV. Finally, Section V concludes the paper.

\section{System Model}

In this section, we introduce the mathematical model of a general GOMIMO system equipped with $N_t$ LEDs and $N_r$ photo-detectors (PDs). The channel model is first described and then the basic principle of GOMIMO is further reviewed.

\subsection{Channel Model}

Letting $\bf{x}$ $= [x_1, x_2, \cdots, x_{N_t}]^T$ be the transmitted signal vector, $\bf{H}$ represent the $N_r \times N_t$ MIMO channel matrix and $\bf{n}$ $= [n_1, n_2, \cdots, n_{N_r}]^T$ denote the additive noise vector, the received signal vector $\bf{y}$ $= [y_1, y_2, \cdots, y_{N_r}]^T$ is obtained by
  \begin{equation}
  \setlength{\abovedisplayskip}{12pt}
  \setlength{\belowdisplayskip}{12pt}
  \bf{y} = H x + n,
  \label{eqn:y}
  \end{equation}
and the corresponding channel matrix $\bf{H}$ can be expressed by
  \begin{equation}
  \setlength{\abovedisplayskip}{12pt}
  \setlength{\belowdisplayskip}{12pt}
  \bf{H} = \left[\begin{array}{ccc}
  h_{1 1} & \cdots & h_{1 N_t}\\
  \vdots & \ddots & \vdots\\
  h_{N_r 1} & \cdots & h_{N_r N_t}
  \end{array}
  \right],
  \label{eqn:H}
  \end{equation}
where $h_{r t}$ ($r = 1, 2, \cdots, N_r; t = 1, 2, \cdots, N_t$) denotes the direct current (DC) channel gain between the $r$-th PD and the $t$-th LED. Assuming that each LED follows the general Lambertian radiation pattern and only the line-of-sight (LOS) transmission is considered, $h_{r t}$ is calculated by \cite{komine2004fundamental} 
  \begin{equation}
  \setlength{\abovedisplayskip}{12pt}
  \setlength{\belowdisplayskip}{12pt}
  h_{r t} = \frac{(l + 1) \rho A}{2\pi d_{r t}^2} \mathrm{cos}^m(\varphi_{r t}) T_s(\theta_{r t}) g(\theta_{r t})   \mathrm{cos}(\theta_{r t}).	
  \label{eqn:h}
  \end{equation}
In (\ref{eqn:h}), $l = -\mathrm{ln}2/\mathrm{ln}(\mathrm{cos}(\Psi))$ denotes the Lambertian emission order, with $\Psi$ being the semi-angle at half power of the LED; $\rho$ and $A$ represent the responsivity and the physical area of the PD, respectively; $d_{r t}$ is the distance between the $r$-th PD and the $t$-th LED; $\varphi_{r t}$ and $\theta_{r t}$ are the emission angle and the incident angle, respectively; $T_s(\theta_{r t})$ is the gain of optical filter; $g(\theta_{r t}) = \frac{n^2}{\mathrm{sin}^2\Phi}$ is the gain of optical lens, where $n$ and $\Phi$ are the refractive index and the half-angle field-of-view (FOV) of the optical lens, respectively. 

Moreover, the additive noise in typical OWC systems consists of both shot and thermal noises, and it is reasonable to model the additive noise as a real-valued zero-mean additive white Gaussian noise (AWGN) \cite{fath2013performance}. Letting $N_0$ denote the noise power spectral density (PSD) and $B$ be the signal bandwidth, the power of the additive noise is given by $P_n = N_0 B$.

\subsection{Principle of GOMIMO}

The concept of GOMIMO was first proposed in \cite{chen2021ofdm}, which aims to fully explore the potential of MIMO transmission for spectral efficiency enhancement of bandlimited OWC systems. Specifically, GOMIMO techniques can be generally divided into two main categories: one is GOSM where all the activated LED transmitters transmit the same signal, and the other is GOSMP where the activated LED transmitters transmit different signals. For more details about the GOMIMO techniques, please refer to our previous work \cite{chen2021ofdm}.

Fig. \ref{fig:diagram} illustrates the block diagram of a general $N_r \times N_t$ GOMIMO system, where $N_a$ ($1 \le N_a \le N_t$) LEDs are activated for signal transmission during GOMIMO mapping. As we can see, the input bits are first divided into two streams: one is fed into the constellation mapper which converts the binary bits into constellation symbols, and the other is sent into the LED index selector which selects the desired LEDs to transmit the generated constellation symbols accordingly. Based on the obtained constellation symbol vector $\bf{c}$ and spatial index vector $\bf{v}$, GOMIMO (GOSM or GOSMP) mapping is performed to generate the transmitted signal vector $\bf{x}$. The mapping tables for GOSM and GOSMP with $N_t = 4$ and $N_a = 2$ are given in insets (a) and (b) of Fig. \ref{fig:diagram}, respectively. At the receiver side, the received signal vector $\bf{y}$ is fed into the GOMIMO detector which finally yields the output bits. The detailed GOMIMO detection schemes will be discussed in the following section.

In typical LED-based OWC systems, intensity modulation with direct detection (IM/DD) is generally applied due to the non-coherence nature of LEDs. As a result, only real-valued non-negative signals can be successfully transmitted in the IM/DD OWC systems \cite{komine2004fundamental}. In this work, unipolar $M$-ary pulse amplitude modulation ($M$-PAM) is adopted as the modulation format for GOMIMO systems. In order to avoid the loss of spatial information when performing GOMIMO mapping, the $M$-PAM symbols cannot have zero values \cite{chen2021ofdm}. Therefore, unipolar non-zero $M$-PAM modulation is utilized here and the corresponding intensity levels are given by
\begin{equation}
  \setlength{\abovedisplayskip}{12pt}
  \setlength{\belowdisplayskip}{12pt}
I_{m}=\frac{2 I_{\textrm{av}}}{M+1} m,~~~~m = 1, \cdots, M,
\end{equation}
where $I_{\textrm{av}}$ denotes the average optical power emitted \cite{fath2013performance}. Using $M$-PAM modulation, the spectral efficiencies (bits/s/Hz) of the $N_r \times N_t$ GOMIMO system with $N_a$ activated LEDs applying GOSM and GOSMP mappings are respectively given by
\begin{equation}
{\eta}_{\textrm{GOSM}} = \log _2 (M) + \lfloor \log _2 (C(N_{t}, N_{a})) \rfloor,
\end{equation}
\begin{equation}
{\eta}_{\textrm{GOSMP}} =N_a \log _2 (M) + \lfloor \log _ 2 (C(N_{t}, N_{a})) \rfloor,
\end{equation}
where $\lfloor \cdot \rfloor$ denotes the floor operator which outputs an integer smaller or equal to its input value and $C(\cdot, \cdot)$ represents the
binomial coefficient.

\section{Detection Schemes for GOMIMO Systems}

In this section, we first introduce two conventional detection schemes for GOMIMO systems utilizing $M$-PAM modulation, including the optimal joint ML detection and the ZF-ML detection. After that, we further present two deep learning-aided detection schemes, including the CSI-based ZF-DNN detection and our newly proposed CSI-free blind DNN detection.

\subsection{Joint ML Detection}

Assuming perfect CSI, joint ML detection is the optimal detection scheme for GOMIMO systems with $M$-PAM modulation. More specifically, the joint ML detector estimates the transmitted constellation and spatial information simultaneously in a joint manner. By applying the joint ML detector, the transmitted signal vector $\mathbf{x}$ can be estimated by
\begin{equation}
  \setlength{\abovedisplayskip}{12pt}
  \setlength{\belowdisplayskip}{12pt}
 \hat{\mathbf{x}}_{\textrm{JML}} = \arg \min_{\mathbf{x} \in \mathbb{X}} {\left\| \mathbf{y} - \mathbf{H} \mathbf{x} \right\|}^{\textrm{2}},
\end{equation}
where ${\lVert \cdot \rVert}_2$ denotes the modulus operator and $\mathbb{X}$ represents the set of all the considered transmitted signal vectors.

Although the joint ML detection can achieve optimal performance, it suffers from high computational complexity. Therefore, it is usually not feasible to apply the joint ML detector in practical GOMIMO systems.

\subsection{ZF-ML Detection}

In order to avoid the high computational complexity of joint ML detection, a low-complexity ZF-ML detection scheme can be applied in GOMIMO systems, which is basically a three-step detection scheme \cite{tavakkolnia2018ofdm,chen2021ofdm}. In the first step, ZF equalization is performed for MIMO demultiplexing. The estimate of the transmitted signal vector $\mathbf{x}$ after ZF equalization can be obtained by 
  \begin{equation}
  \setlength{\abovedisplayskip}{12pt}
  \setlength{\belowdisplayskip}{12pt}
  \bf{\hat{x}}_{\textrm{ZF}} = H^{\dag} y = x + H^{\dag} n,	
  \label{eqn:x_hat}
  \end{equation}
where $\bf{H^{\dag}}$ denotes the pseudo inverse of $\bf{H}$ \cite{chen2020user}. 

In the second step, ML detection is executed to obtain the estimate of the spatial index vector according to $\bf{\hat{x}}_{\textrm{ZF}}$. Finally, in the third step, the estimate of the constellation symbol vector can be obtained accordingly by using $\bf{\hat{x}}_{\textrm{ZF}}$ and the estimate of the spatial index vector. For more details about the principle of ZF-ML detection for GOMIMO systems, please refer to our previous work \cite{chen2021ofdm}.

Compared with joint ML detection, the computational complexity of ZF-ML detection is significantly reduced. Nevertheless, the performance of ZF-ML detection is also largely degraded in comparison to that of joint ML detection, which can be explained as follows. On the one hand, ZF equalization inevitably causes severe noise amplification due to the high channel correlation in typical MIMO-OWC systems \cite{fath2013performance}, which might greatly degrade the performance of GOMIMO systems. On the other hand, the detection error of spatial symbols might propagate to the estimation of the constellation symbols \cite{chen2021deep}, which leads to further substantial performance degradation of GOMIMO systems.

\begin{figure*}[!t]
\centering
{\includegraphics[width=1.8\columnwidth]{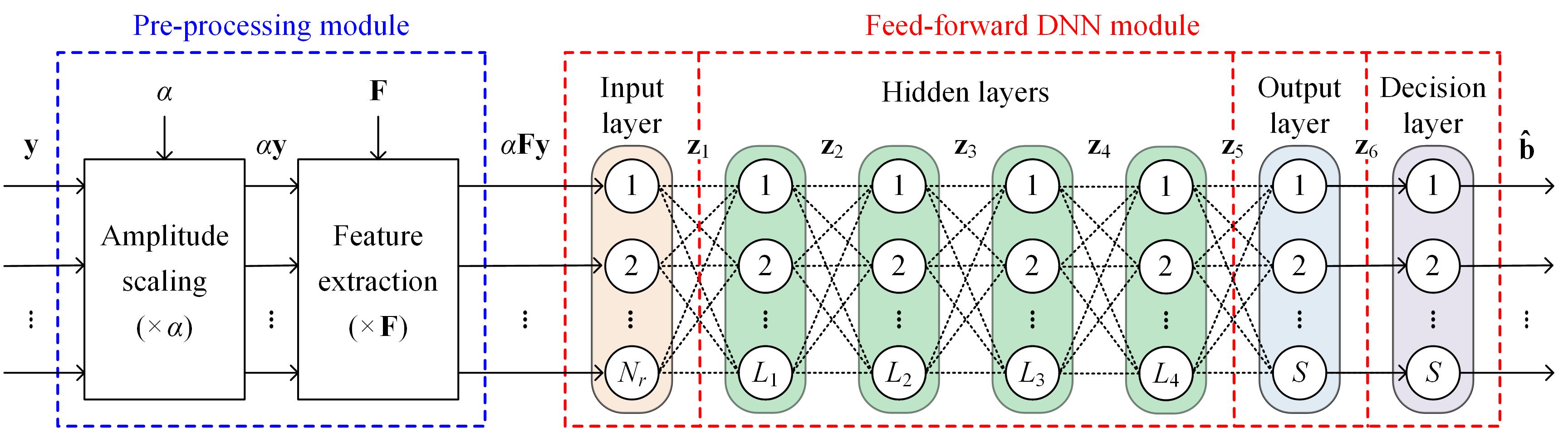}} 
\caption{Schematic diagram of the proposed CSI-free blind DNN detector consisting of a pre-processing module and a feed-forward DNN module.} 
\label{fig:schematic}
\end{figure*}

\subsection{CSI-Based ZF-DNN Detection}

To efficiently address both the high computational complexity issue of joint ML detection and the noise amplification and error propagation issues of ZF-ML detection, a ZF-DNN detection has been proposed for GOSMP systems in \cite{wang2020deep}. The key idea of the ZF-DNN detection scheme is to employ a feed-forward DNN module to directly and simultaneously estimate the transmitted spatial and constellation bits by taking the ZF equalized signal vector $\bf{\hat{x}}_{\textrm{ZF}}$ as input. For more details about the implementation of the ZF-DNN detector, please refer to \cite{wang2020deep,chen2021deep}. In a word, the feed-forward DNN module can fulfill the tasks of spatial index vector estimation, constellation symbol vector estimation, spatial symbol demodulation and constellation symbol demodulation at the same time. Simulation results in \cite{wang2020deep} clearly show that, by selecting a proper training signal-to-noise ratio (SNR), the ZF-DNN detector can achieve nearly the same BER performance as the optimal joint ML detector, but with a significantly reduced computational complexity.

Despite the near-optimal BER performance and low computational complexity of the ZF-DNN detector, it takes the ZF equalized signal vector $\bf{\hat{x}}_{\textrm{ZF}}$ as the input of the feed-forward DNN module. As per (\ref{eqn:x_hat}), $\bf{\hat{x}}_{\textrm{ZF}}$ is obtained by multiplying the received signal vector $\bf{y}$ with $\bf{H^{\dag}}$, i.e., the  pseudo inverse of the channel matrix $\bf{H}$. Generally, the CSI (i.e., the channel matrix) can be efficiently estimated by transmitting training symbols \cite{wang2014demo}. Nevertheless, the use of training symbols for accurate CSI estimation inevitably reduces the achievable data rate of GOMIMO systems, especially for low SNR scenarios. Furthermore, since the channel matrix is highly related to the specific location of the MIMO receiver, i.e., the PD array, channel estimation needs to be executed instantaneously with the change of receiver location. In consequence, instantaneous channel estimation inevitably introduces additional communication time delay and computational complexity in practical GOMIMO systems. 

\subsection{Proposed CSI-Free Blind DNN Detection}

Considering the many disadvantages of CSI-based ZF-DNN detection due to the requirement of instantaneous CSI for ZF equalization, in this work, we for the first time propose a novel CSI-free blind DNN detection scheme for GOMIMO systems. Fig. \ref{fig:schematic} depicts the schematic diagram of the proposed CSI-free blind DNN detector, which consists of a pre-processing module and a feed-forward DNN module. It can be seen that a pre-processing module is placed in front of the feed-forward DNN module in the proposed CSI-free blind DNN detector, which is the key to deal with the impact of MIMO transmission through free-space channels and hence realize blind detection without the need of CSI. Specifically, as can be found from (\ref{eqn:y}), the impact of MIMO transmission on the transmitted signal vector $\bf{x}$ can be characterized from the following two aspects. Firstly, since the channel coefficients in typical MIMO-OWC systems are within the region from $10^{-6}$ to $10^{-4}$ \cite{fath2013performance,ying2015joint}, the electrical path loss caused by MIMO transmission is about 80 to 120 dB. Secondly, MIMO transmission also inevitably leads to channel crosstalk, which might cause severe ICI, especially for GOSMP systems. As a result, the designed pre-processing module should be able to address both the path loss issue and the channel crosstalk issue caused by MIMO transmission. 

As shown in Fig. \ref{fig:schematic}, our specially designed pre-processing module mainly contains two parts: one is the amplitude scaling part and the other is the feature extraction part. Specifically, the amplitude scaling part is adopted to address the path loss issue by multiplying the received signal vector $\bf{y}$ with a scaling factor $\alpha$. Note that a proper $\alpha$ value is determined in advance for each receiver location in the GOMIMO system, and hence no instantaneous CSI is needed to achieve amplitude scaling. Moreover, the feature extraction part is used to address the channel crosstalk issue, which multiplies the scaled received signal vector $\alpha \bf{y}$ by a feature matrix $\bf{F}$. Hence, the output signal vector of the pre-processing module in the CSI-free blind DNN detector, i.e., $\bf\hat{y}$ $= [\hat{y}_1, \hat{y}_2, \cdots, \hat{y}_{N_r}]^T$, can be obtained by
 \begin{equation}
 \setlength{\abovedisplayskip}{12pt}
 \setlength{\belowdisplayskip}{12pt}
 \bf\hat{y} = \alpha\bf{F}\bf{y}.
 \label{eqn:y2}
 \end{equation}

In order to provide enough information for the following feed-forward DNN module to efficiently learn and remove the impact of channel crosstalk caused by MIMO transmission, the feature matrix $\bf{F}$ should be able to reflect all the potential signal superposition cases at the receiver side. Consequently, according to the mapping tables of both GOSM and GOSMP in Fig. \ref{fig:diagram}, we adopt the corresponding unified mapping matrix as the feature matrix, i.e.,
\begin{equation}
\bf{F}=\left[\begin{array}{llll}
1 & 1 & 0 & 0 \\
1 & 0 & 1 & 0 \\
0 & 1 & 0 & 1 \\
0 & 0 & 1 & 1
\end{array}\right].
\end{equation}

\begin{table}[!t]
\centering
\caption{Simulation Parameters}\label{tab:Parameters}
\begin{tabular}{c|c}
\toprule
Parameter & Value\\
\midrule
Room dimension & $5 \ \textrm{m} \times 5 \ \textrm{m} \times 3 \ \textrm{m}$\\
Height of receiving plane & 0.85 m\\
Number of LEDs & 4\\
Semi-angle at half power of LED & $60^{\circ}$\\
LED spacing & 2.5 m\\
Gain of optical filter & 0.9\\
Refractive index of optical lens & 1.5\\
Half-angle FOV of optical lens & $72^{\circ}$\\
Number of PDs & 4\\
Responsivity of PD & 1 $\mathrm{A/W}$\\
Active area of PD & 1 $\textrm{cm}^2$\\
PD spacing & 10 cm\\
Number of activated LEDs, $N_a$& 2\\
PAM levels, $M$& 4\\
\bottomrule
\end{tabular}
\end{table}

Subsequently, the pre-processed signal vector $\bf\hat{y}$ is fed into a feed-forward DNN module, which mainly consists of an input layer, multiple hidden layers, an output layer and a decision layer. Since $\bf\hat{y}$ is a vector with $N_r$ elements, the input layer contains $N_r$ neurons accordingly. Moreover, we set totally four fully-connected hidden layers in the feed-forward DNN module, which are used to learn the statistical characteristics of both the input signal and the additive noise. The number of neurons in the $i$-th ($1 \le i \le 4$) hidden layer is denoted by $L_i$, and the rectified linear unit (ReLU) function, i.e., $f_{\textrm{ReLU}}(\alpha)=\max (0, \alpha)$, is adopted as the activation function of the hidden layers. For the output layer, it adopts the Sigmoid function, i.e., $f_{\textrm{Sigmoid }}(\alpha)=1 /\left(1+\exp ^{-\alpha}\right)$, as the activation function to generate a fuzzy bit information, so as to map the output of each neuron within the range [0, 1]. Since the DNN detector takes the input binary bits corresponding to a transmitted signal vector as the output, both the output layer and the decision layer have the same number of neurons, which is equal to the spectral efficiency of the GOMIMO system, i.e., $S = \eta_{\textrm{GOMIMO}}$. Therefore, letting $\mathbf{z}_k$ denote the output of the $k$-th ($1 \le k \le 6$) layer of the feed-forward DNN module, the corresponding input-output relationship can be described by
\begin{equation}
{\mathbf{z}}_k = \left\{\begin{array}{ll}
\alpha\bf{F}\bf{y}, & k = 1\\
f_{\textrm{ReLU}}\left(\mathbf{W}_{k-1} {\mathbf{z}}_{k-1}+\mathbf{b}_{k-1}\right), & 2 \leq k \leq 5\\
f_{\textrm{Sigmoid}}\left(\mathbf{W}_{k-1} {\mathbf{z}}_{k-1}+\mathbf{b}_{k-1}\right), & k=6
\end{array},\right.
\end{equation}
where $\mathbf{W}_p$ and $\mathbf{b}_p$ with $1 \le p \le 5$ represent the corresponding weight matrix and the bias vector, respectively. 

Finally, the decision layer is utilized to determine the fuzzy output of each neuron in the output layer to be 0 or 1. Letting $\mathbf{z}_6 = [z_1, z_2, \dots z_S]^T$ and $\bf\hat{b}$ $= [\hat{b}_1, \hat{b}_2, \cdots, \hat{b}_S]^T$ respectively denote the fuzzy output vector of the output layer and the final output binary bit vector, the $q$-th ($q = 1, 2, \cdots, S$) binary bit in $\bf\hat{b}$ can be estimated by
\begin{equation}
\hat{b}_q = \left\{\begin{array}{ll}
0, &  z_q < 0.5
\\
1, &  z_q \geq 0.5
\end{array}.\right.
\end{equation}

In the proposed CSI-free blind DNN detector, we adopt the mean-square error (MSE) loss function to measure the difference between the transmitted bit vector $\bf{b}$ and the corresponding estimated bit vector $\bf\hat{b}$, which is given by
\begin{equation}
e_{\textrm{MSE}} = \frac{1}{S} \left\| \bf\hat{b} - \bf{b} \right\| ^ 2.
\end{equation}

\section{Simulation Results}

In this section, we evaluate and compare the performance of four different detection schemes in a typical indoor GOMIMO system through numerical simulations. 

\begin{table}[!t]
\caption{Parameters of the Blind DNN Detector for GOSM}
\centering
\begin{tabular}{c|c}
\toprule
Parameter & Value\\
\midrule
Receiver locations & (2.5 m, 2.5 m, 0.85) $\mid$ (0 m, 0 m, 0.85)\\
Number of input nodes & 4 \\
Number of hidden layers & 4 \\
Number of neurons & $L_1$ = 128, $L_2$ = 64, $L_3$ = 32, $L_4$ = 16\\
Number of output nodes & 4 \\
Hidden layer activation & ReLU \\
Output layer activation & Sigmoid \\
Loss function & MSE \\
Optimizer & Adamax \\
Learning rate & 0.01 $\mid$ 0.001 \\
Length of training set & 150000 \\
Length of validation set & 50000 \\
Scaling factor & $1 \times 10^5$ $\mid$ $2 \times 10^5$ \\
\bottomrule
\end{tabular}
\label{tableGSM}
\end{table}

\begin{table}[!t]
\caption{Parameters of the Blind DNN Detector for GOSMP}
\centering
\begin{tabular}{c|c}
\toprule
Parameter & Value\\
\midrule 
Receiver locations & (2.5 m, 2.5 m, 0.85) $\mid$ (0 m, 0 m, 0.85)\\
Number of input nodes & 4 \\
Number of hidden layers & 4 \\
Number of neurons & $L_1$ = 64, $L_2$ = 64, $L_3$ = 64, $L_4$ = 64\\
Number of output nodes & 6\\
Hidden layer activation & ReLU  \\
Output layer activation & Sigmoid \\
Loss function & MSE \\
Optimizer & Adamax \\
Learning rate & 0.01 $\mid$ 0.005 \\
Length of training set & 150000 \\
Length of validation set & 50000\\
Scaling factor & $1 \times 10^5$ $\mid$ $1 \times 10^6$ \\
\bottomrule
\end{tabular}
\label{tableGSMP}
\end{table}

\subsection{Simulation Setup}

In our simulations, we consider a 4 $\times$ 4 ($N_r$ = $N_t$ = 4) GOMIMO system configured in a typical 5 m $\times$ 5 m $\times$ 3 m room. The 2 $\times$ 2 square LED array is placed at the center of the ceiling and the spacing between two adjacent LEDs is 2 m. The height of the receiving plane is 0.85 m, and two receiver locations over the receiving plane, i.e., the center (2.5 m, 2.5 m, 0.85 m) and the corner (0 m, 0 m, 0.85 m), are considered for performance evaluation. The receiver consists of a 2 $\times$ 2 square PD array, where the spacing between two adjacent PDs is 10 cm. For both GOSM and GOSMP mappings, two out of four LEDs are activated for signal transmission, i.e., $N_a$ = 2. Moreover, unipolar non-zero 4-PAM modulation is adopted in the GOMIMO system, and hence the corresponding spectral efficiencies for GOSM and GOSMP mappings are 4 and 6 bits/s/Hz, respectively. In addition, we adopt transmitted SNR as the measure to evaluate the BER performance of the GOMIMO system \cite{fath2013performance,chen2021ofdm}. The other simulation parameters of the GOMIMO system can be found in Table \ref{tab:Parameters}. 

\begin{figure}[!t]
\centering
\subfigure[GOSM]{
\label{fig:MSEGSM}
\includegraphics[width=0.9\columnwidth]{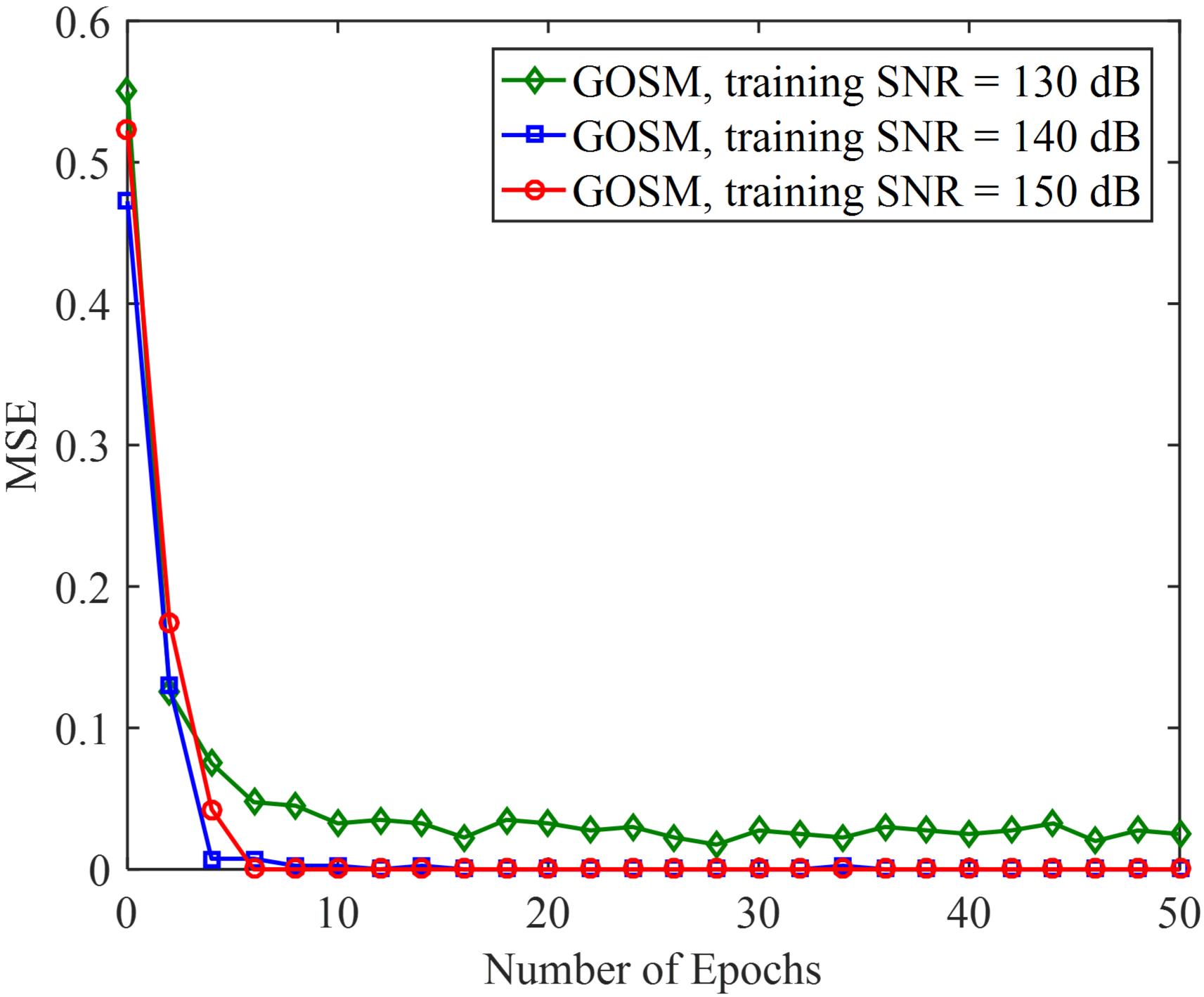}}
\subfigure[GOSMP]{
\label{fig:MSEGSMP}
\includegraphics[width=0.9\columnwidth]{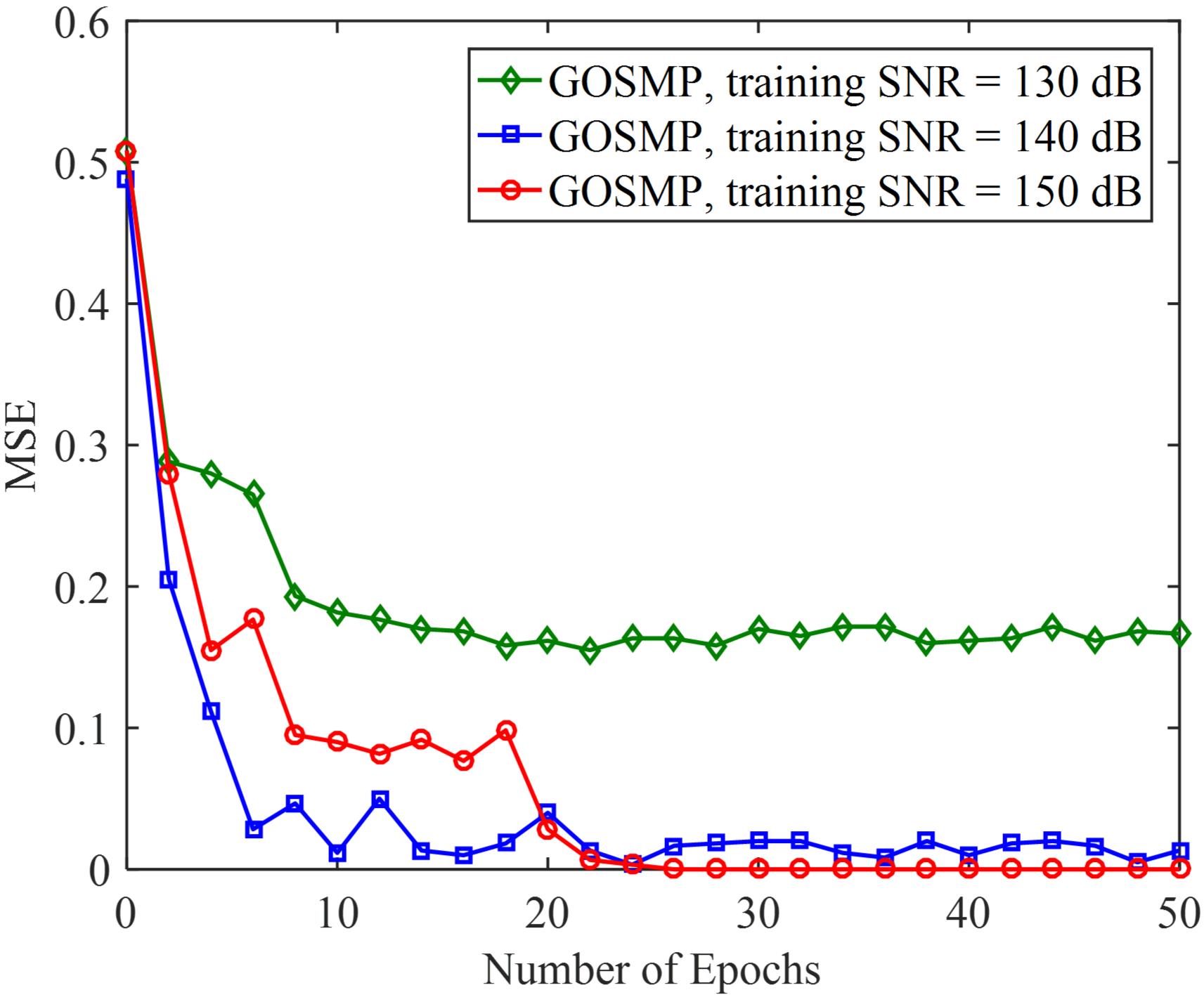}}
\caption{MSE loss of the proposed CSI-free blind DNN detector with receiver located at the center of the receiving plane for (a) GSM and (b) GSMP.}
\label{fig:MSE}
\end{figure}

The detailed parameters of the CSI-free blind DNN detectors for GOSM and GOSMP are given in Tables II and III, respectively. For GOSM, the number of neurons of four hidden layers is 128, 64, 32 and 16, respectively. The learning rate is 0.01 when the receiver is located at the center of the receiving plane, while it is reduced to 0.001 when the receiver is moved to the corner. Moreover, the scaling factors are set to $1 \times 10^5$ and $2 \times 10^5$ when the receiver is located at the center and the corner, respectively. For GOSMP, every hidden layer contains 64 neurons, and the learning rates are 0.01 and 0.005 when the receiver is located at the center and the corner of the receiving plane, respectively. In addition, the scaling factors of $1 \times 10^5$ and $1 \times 10^6$ are used for center and corner received locations, respectively. For both GOSM and GOSMP, the lengths of training set and validation set are assumed to be 150000 and 50000, respectively. In order to accelerate the convergence speed, we use the mini-batch technique in training and each mini-batch contains 100 transmitted signal vectors.

\subsection{MSE Loss}

We first analyze the MSE loss of the proposed CSI-free blind DNN detector in the 4 $\times$ 4 GOMIMO system. Figs. \ref{fig:MSE}(a) and (b) show the MSE losses versus the number of epochs for GOSM and GOSMP, respectively, where the receiver is located at the center of the receiving plane. As we can see, the MSE loss decreases rapidly with the increase of training epochs for both GOSM and GOSMP. Moreover, the MSE loss is much reduced when a higher training SNR is used, especially for GOSMP. It can be seen from Fig. \ref{fig:MSE}(a) that the MSE loss for GOSM fast converges with only a few epochs. For GOSMP, as shown in Fig. \ref{fig:MSE}(b), about 20 epochs are required for the MSE loss to converge. Hence, owing to the use of the mini-batch technique, the CSI-free blind DNN detector only requires a very limited number of epochs for efficient training, indicating that it can be deployed rapidly in practical applications.

\subsection{BER Performance}

\begin{figure}[!t]
\centering
\subfigure[GOSM, center]{
\label{fig:BERGSMcenter}
\includegraphics[width=0.9\columnwidth]{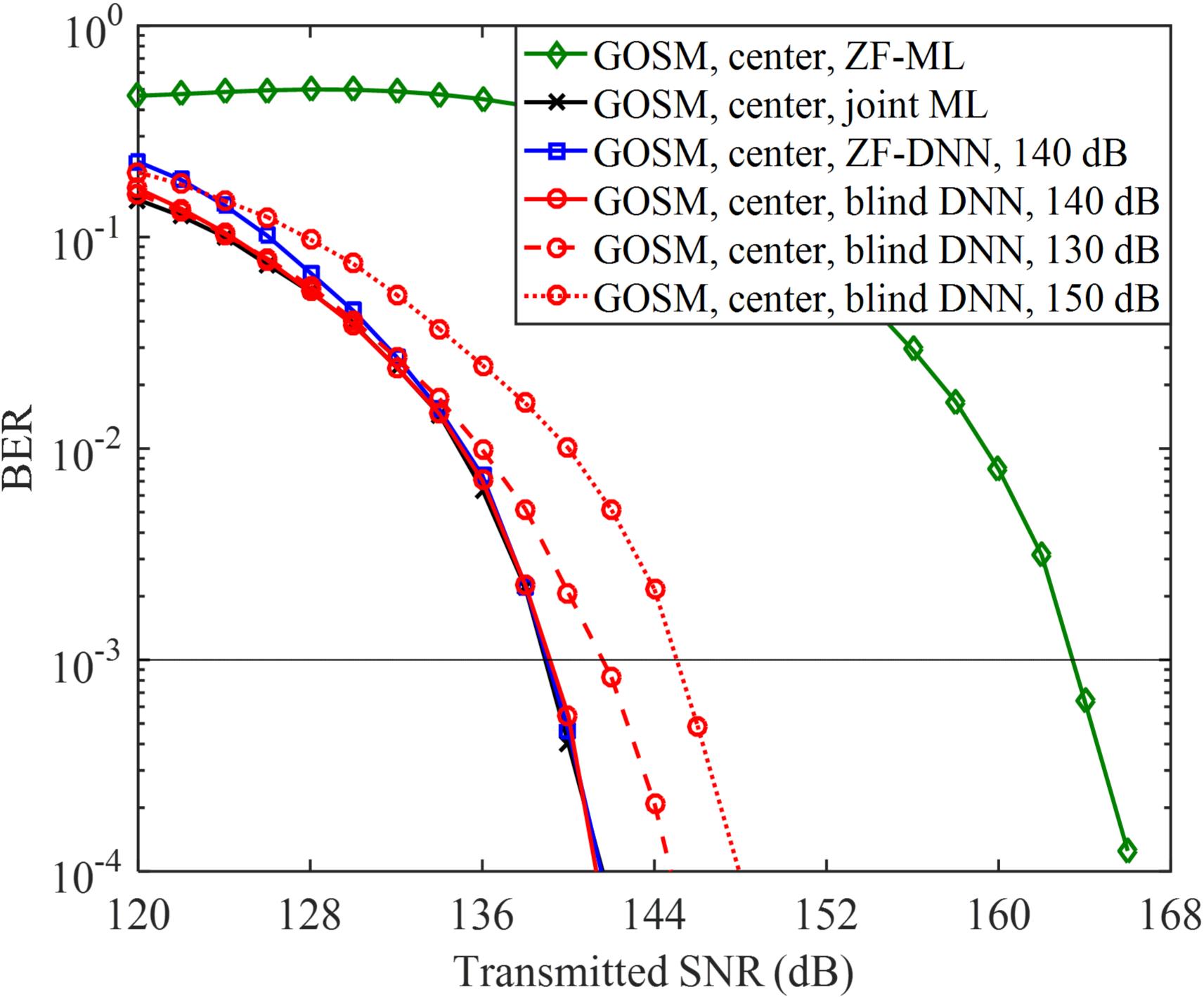}}
\subfigure[GOSM, corner]{
\label{fig:BERGSMcorner}
\includegraphics[width=0.9\columnwidth]{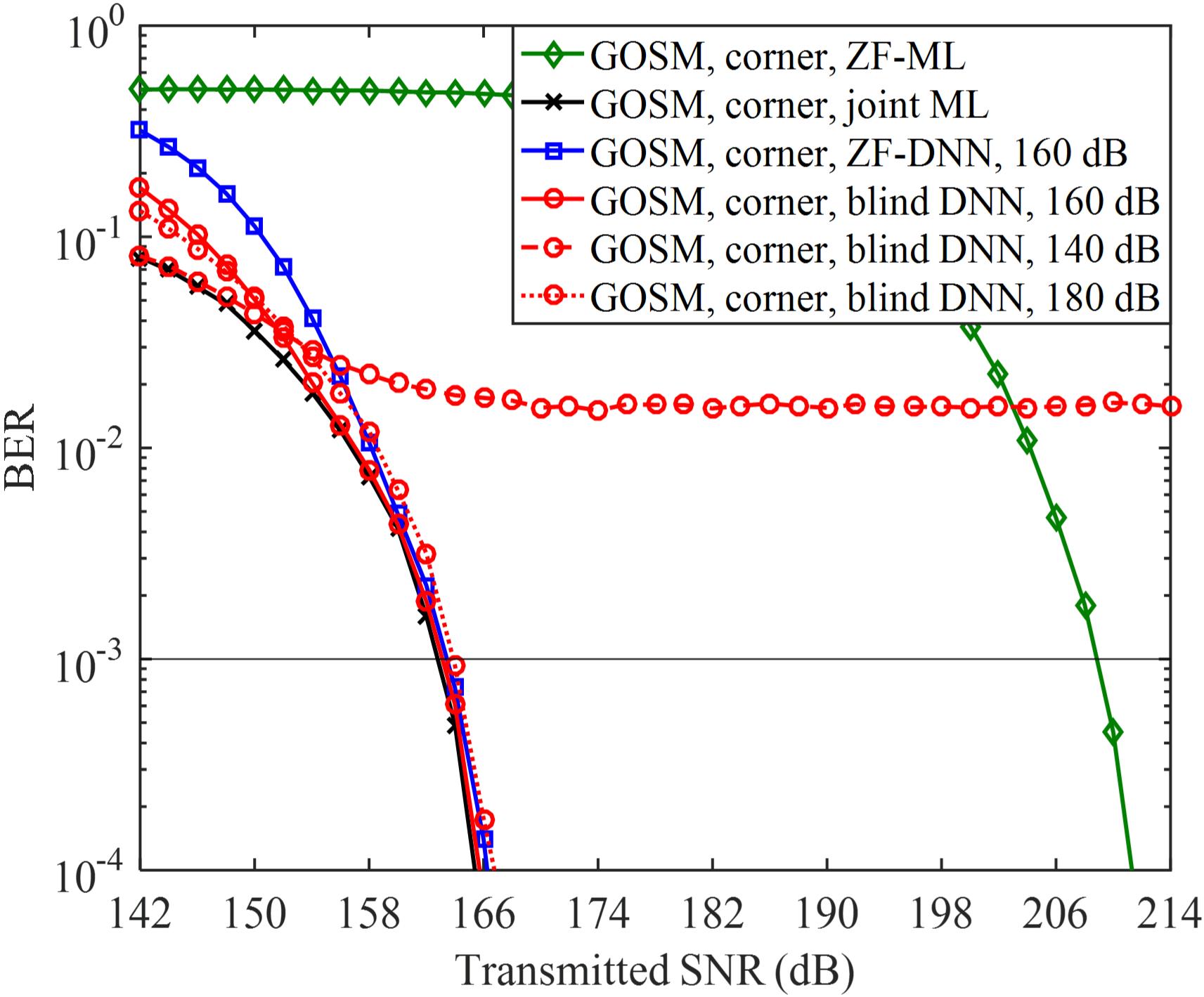}}
\caption{BER comparison of the proposed CSI-free blind DNN detector and three benchmark detectors for GOSM at (a) the center and (b) the corner.}
\label{fig:BERGSM}
\end{figure}

\begin{figure}[!t]
\centering
\subfigure[GOSMP, center]{
\label{fig:BERGSMPcenter}
\includegraphics[width=0.9\columnwidth]{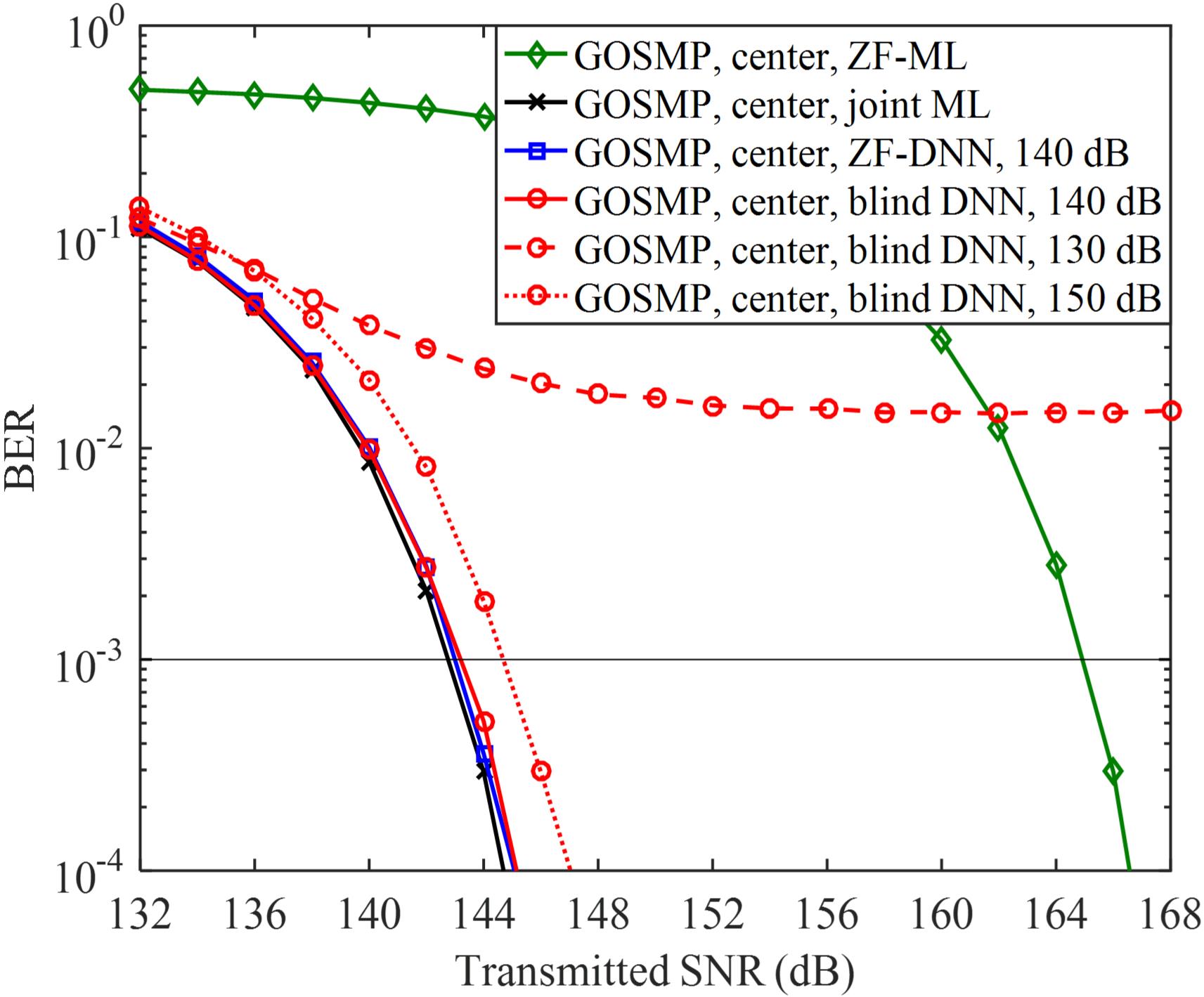}}
\subfigure[GOSMP, corner]{
\label{fig:BERGSMPcorner}
\includegraphics[width=0.9\columnwidth]{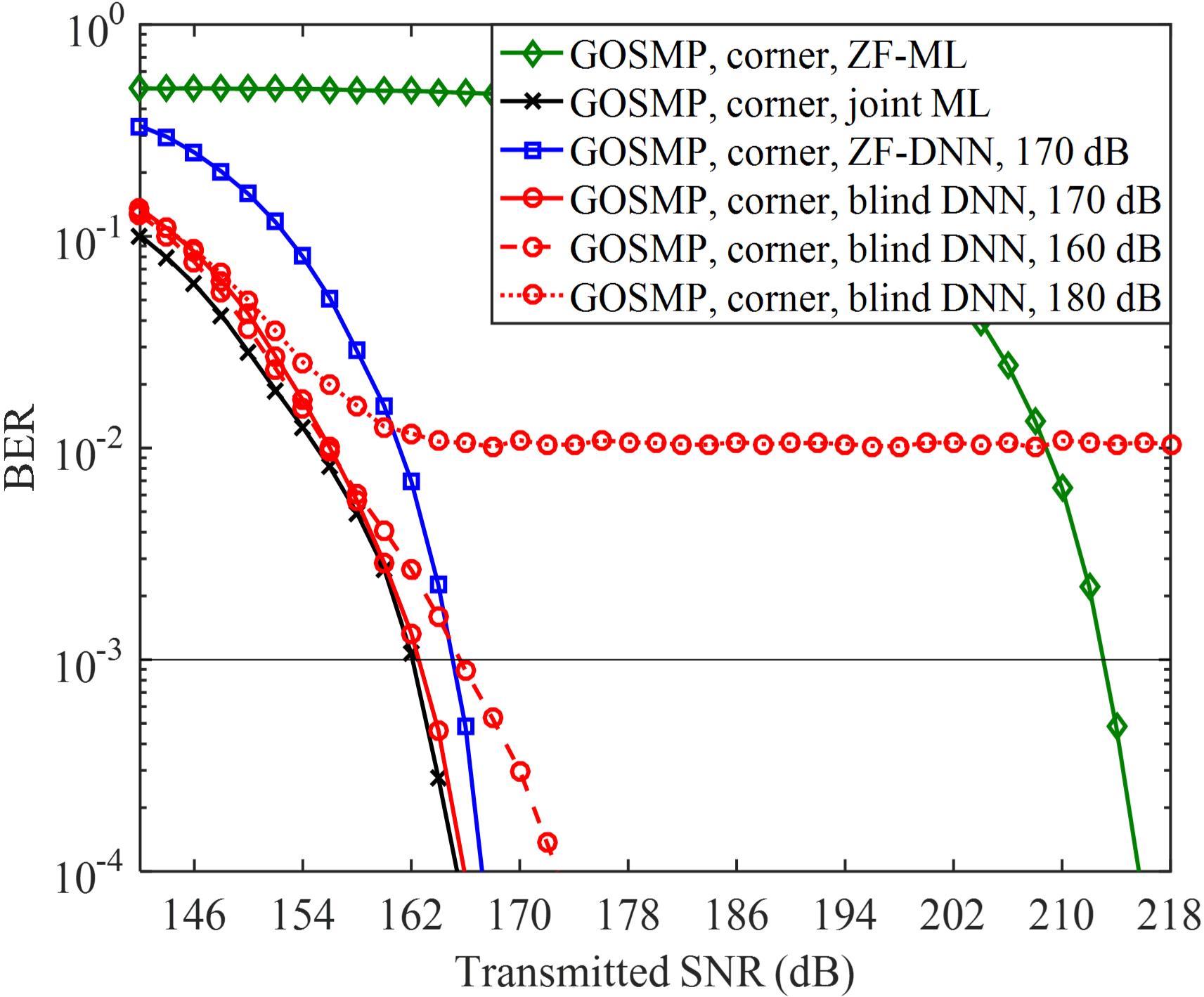}}
\caption{BER comparison of the proposed CSI-free blind DNN detector and three benchmark detectors for GOSMP at (a) the center and (b) the corner.}
\label{fig:BERGSMP}
\end{figure}

We further evaluate and compare the BER performance of the proposed CSI-free blind DNN detector with the other three benchmark detectors in the 4 $\times$ 4 GOMIMO system. Figs. \ref{fig:BERGSM}(a) and (b) compare the BER performance of four detectors for GOSM with the receiver located at the center and the corner of the receiving plane, respectively. When the receiver is located at the center of the receiving plane, as shown in Figs \ref{fig:BERGSM}(a), the ZF-ML detector requires a high transmitted SNR of 163.4 dB to achieve the target BER of $10^{-3}$. However, the required SNR to reach BER = $10^{-3}$ is reduced to 138.9 dB for the joint ML detector. As a result, a substantial 24.5-dB SNR gain can be obtained by the joint ML detector in comparison to the ZF-ML detector, which is mainly because that the ZF-ML detector suffers from severe noise amplification and error propagation. Moreover, it can be further seen that the ZF-DNN detector with an optimal 140-dB training SNR can achieve comparable BER performance as the joint ML detector in the high SNR region, suggesting the excellent error performance of the ZF-DNN detector under the condition of accurate CSI for ZF equalization. Finally, for our proposed CSI-free blind DNN detector, we investigate the impact of training SNR on its error performance and three different training SNRs of 130, 140 and 150 dB are considered. It is clearly shown that the CSI-free blind DNN detector with 140-dB training SNR achieves nearly the same BER performance as the joint ML detector across the whole SNR region, which slightly outperforms the ZF-DNN detector in the low SNR region. However, the joint ML detector outperforms the CSI-free blind DNN detector when a lower training SNR of 130 dB or a higher training SNR of 150 dB is adopted, and the reasons can be explained as follows. The DNN module can better learn the statistics of the noise with a relatively small training SNR, while the statistics of the data symbols can be more accurately learned when the training SNR is relatively large. As a result, there exists an optimal training SNR which can make a trade-off for the DNN module to learn the statistics of both the noise and the data symbols and hence lead to a minimum overall BER. When the receiver is moved to the corner of the receiving plane, as shown in Figs \ref{fig:BERGSM}(b), we can observe that the joint ML detector outperforms the ZF-ML detector by an SNR gain of more than 40 dB at BER = $10^{-3}$, while the ZF-DNN detector with an optimal training SNR of 160 dB obtains near-optimal BER performance as the joint ML detector only for relatively low BERs. Furthermore, the CSI-free blind DNN detector achieves comparable BER performance as the joint ML detector in the high SNR region, which outperforms the ZF-DNN detector in the low SNR region. It should be noted that an error floor occurs for the CSI-free blind DNN detector with a lower training SNR of 140 dB, which is mainly due to the insufficient learning of the statistics of the data symbols under a very noisy environment.

The BER versus transmitted SNR for GOSMP is plotted in Fig. \ref{fig:BERGSMP}. As we can see, the ZF-DNN detector with an optimal training SNR can achieve very close performance as the joint ML detector when the receiver is located at the center of the receiving plane, but it performs worse than the joint ML detector when the receiver is moved to the corner, especially in the low SNR region. In contrast, the proposed CSI-free blind DNN detector can achieve comparable BER performance as the joint ML detector for both center and corner receiver locations. Moreover, error floors occur for the CSI-free blind DNN detector when the adopted training SNR is too small or too large. It can be further observed from Figs. \ref{fig:BERGSM} and \ref{fig:BERGSMP} that the optimal training SNRs for the ZF-DNN detector and the CSI-free blind DNN detector at the same receiver location are generally the same in GOMIMO systems.

\subsection{Impact of Input Pre-processing}

\begin{figure}[!t]
\centering
{\includegraphics[width=.9\columnwidth]{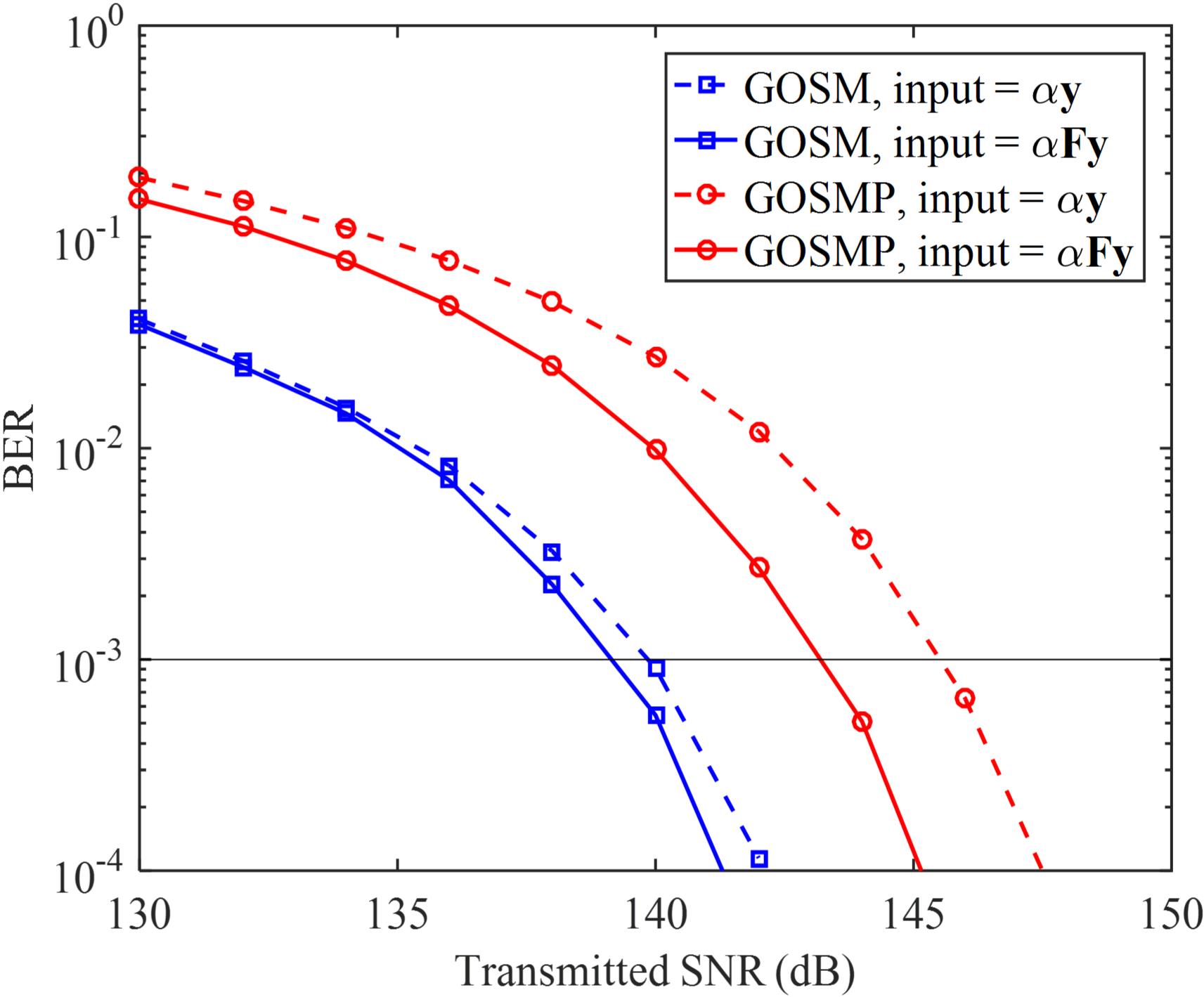}} 
\caption{BER comparison of the proposed CSI-free blind DNN detector where the feed-forward DNN module having different inputs for both GOSM and GOSMP at the center of the receiving plane.} 
\label{fig:input}
\end{figure}

\begin{figure}[!t]
\centering
\subfigure[GOSM]{
\label{fig:BERalphaGSM}
\includegraphics[width=0.9\columnwidth]{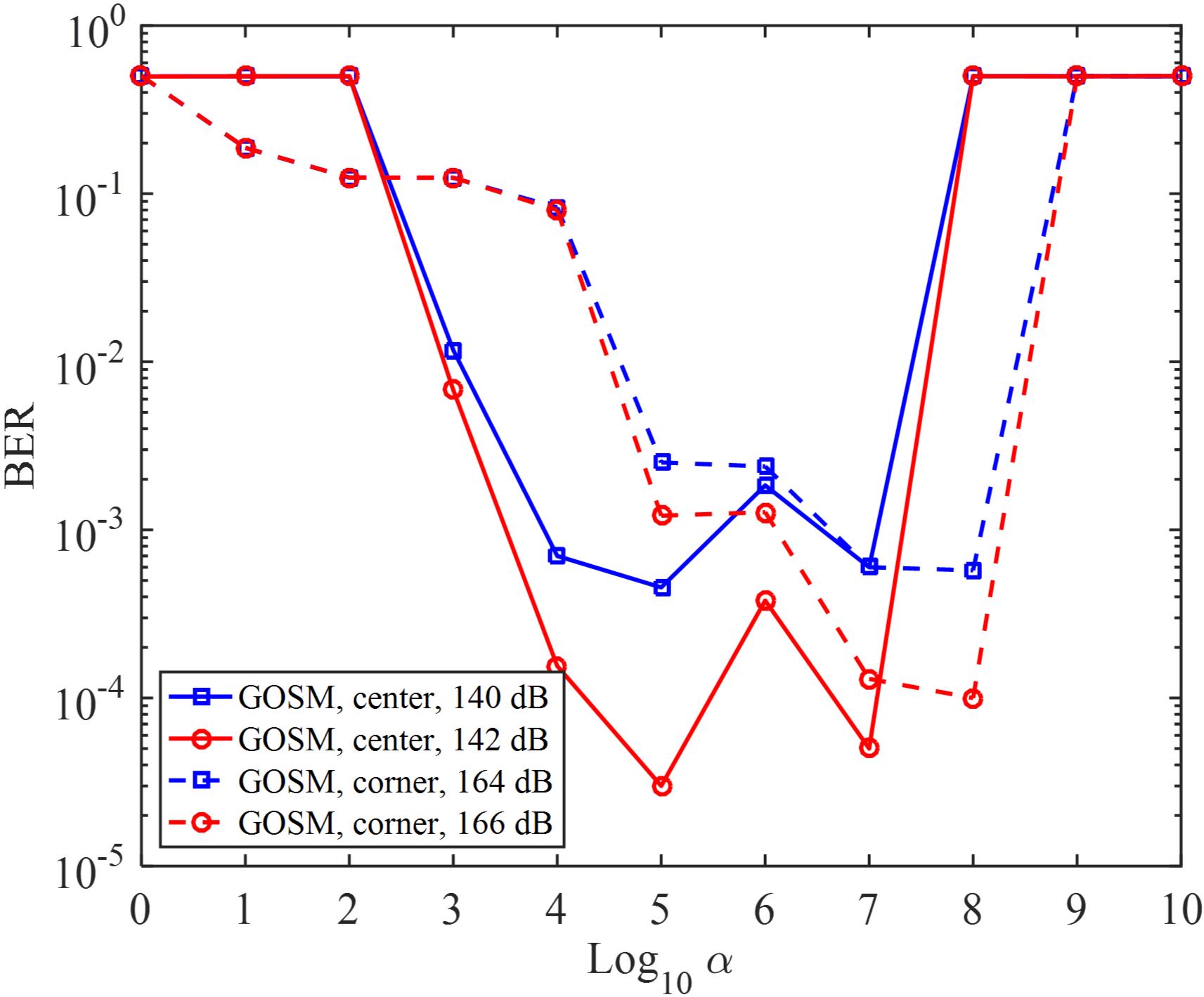}}
\subfigure[GOSMP]{
\label{fig:BERalphaGSMP}
\includegraphics[width=0.9\columnwidth]{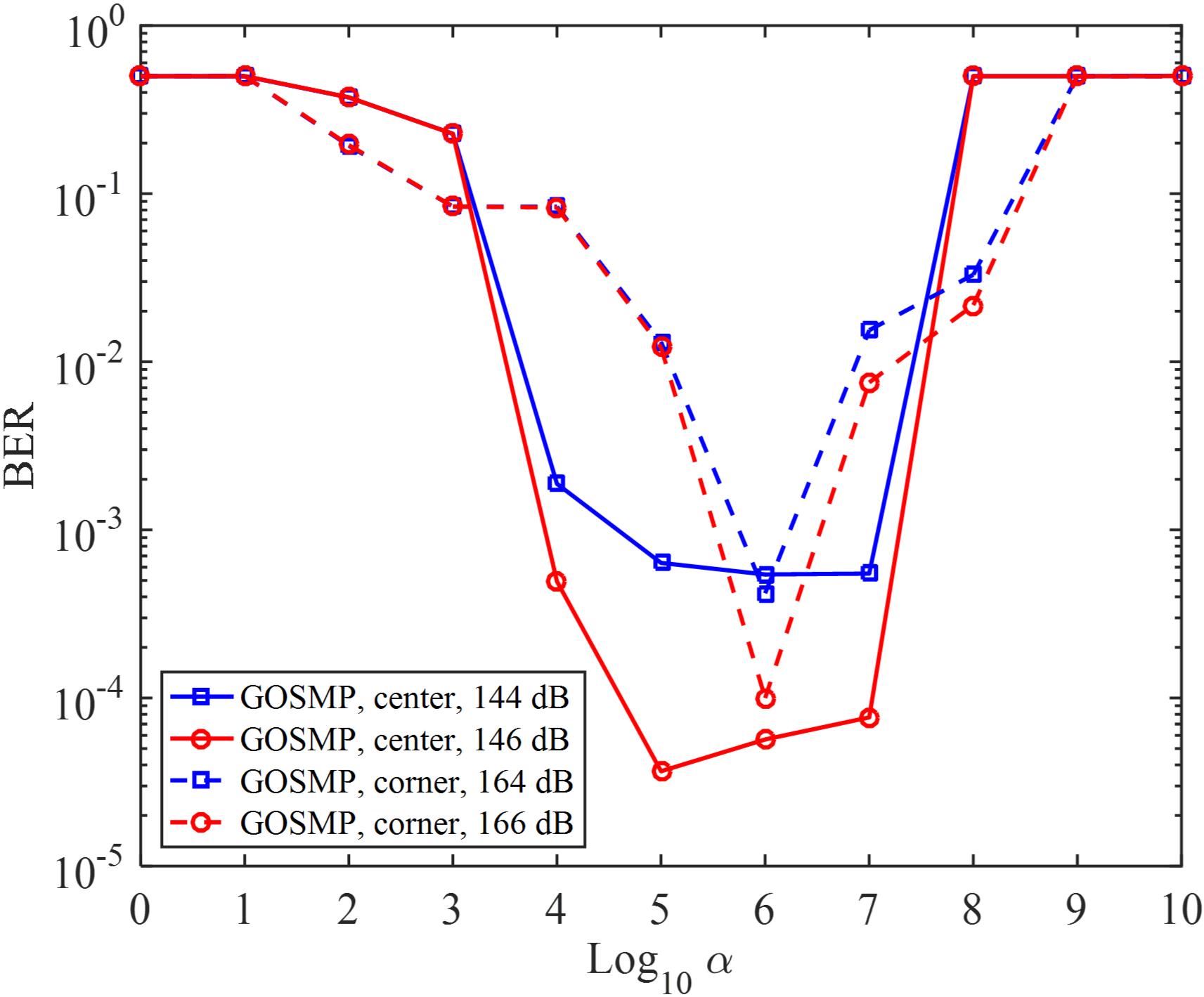}}
\caption{BER vs. $\textrm{log}_{10} \alpha$ of the proposed CSI-free blind DNN detector for (a) GOSM and (b) GOSMP.}
\label{fig:BERalpha}
\end{figure}

It can be seen from Fig. \ref{fig:schematic} that the pre-processing module, which pre-processes the input of the feed-forward DNN module, plays a vital role to guarantee that the proposed CSI-free blind DNN detector can successfully perform MIMO detection blindly without the need of CSI. In the next, we evaluate the impact of input pre-processing on the performance of the CSI-free blind DNN detector. Here, two different inputs of the feed-forward DNN module are considered: one is $\alpha \bf{y}$, i.e., the pre-processing module only performs amplitude scaling, and the other is $\alpha \bf{F y}$, i.e., the pre-processing module performs both amplitude scaling and feature extraction. Fig. \ref{fig:input} compares the BER performance of the proposed CSI-free blind DNN detector where the feed-forward DNN module having different inputs for both GOSM and GOSMP with the receiver located at the center of the receiving plane. As we can see, for GOSM, the BER performance is only slightly improved when the input is changed from $\alpha \bf{y}$ to $\alpha \bf{F y}$, and the SNR gain at BER = $10^{-3}$ is only 0.8 dB. In contrast, for GOSMP, a noticeable BER improvement can be obtained by replacing the input $\alpha \bf{y}$ with $\alpha \bf{F y}$, and the corresponding SNR gain at BER = $10^{-3}$ is increased to 2.4 dB. The difference between BER improvements for GOSM and GOSMP can be explained as follows. As discussed in Section III.D, since the feature matrix $\bf{F}$ contains the spatial mapping information of GOMIMO systems, the feed-forward DNN module can use the spatial mapping information to remove the channel crosstalk. As a result, the feed-forward DNN module with input $\alpha \bf{F y}$ can efficiently mitigate the adverse effect of error propagation. However, in GOSM systems, the activated LEDs are used to transmit the same signal and hence error propagation only leads to reduced diversity gain, which might not significantly degrade the BER performance. In contrast, since the activated LEDs transmit different signals in GOSMP systems, error propagation leads to the missing of constellation information and hence results in significant BER degradation. 

\begin{figure}[!t]
\centering
\subfigure[GOSM]{
\label{fig:TimeGSM}
\includegraphics[width=0.9\columnwidth]{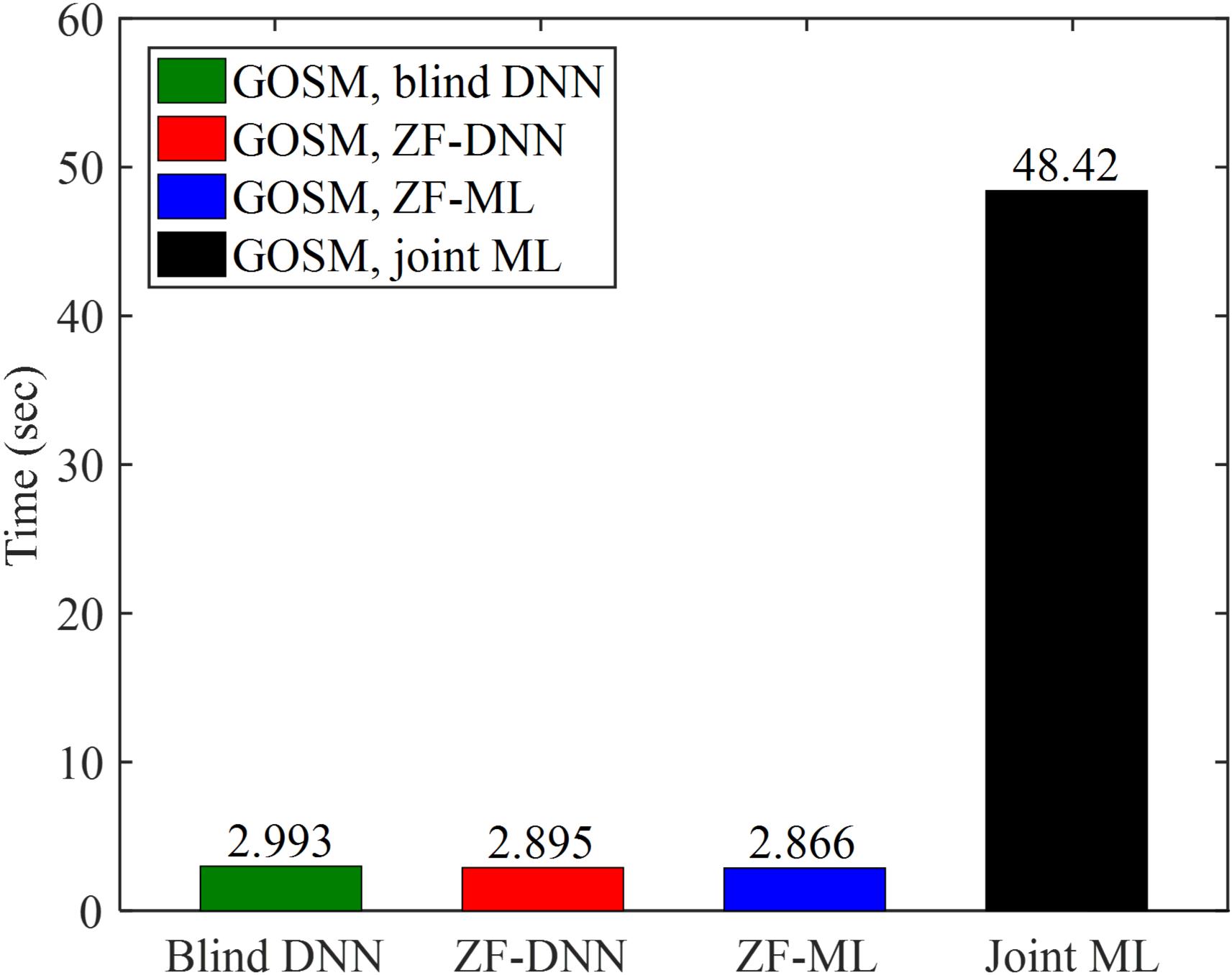}}
\subfigure[GOSMP]{
\label{fig:TimeGSMP}
\includegraphics[width=0.9\columnwidth]{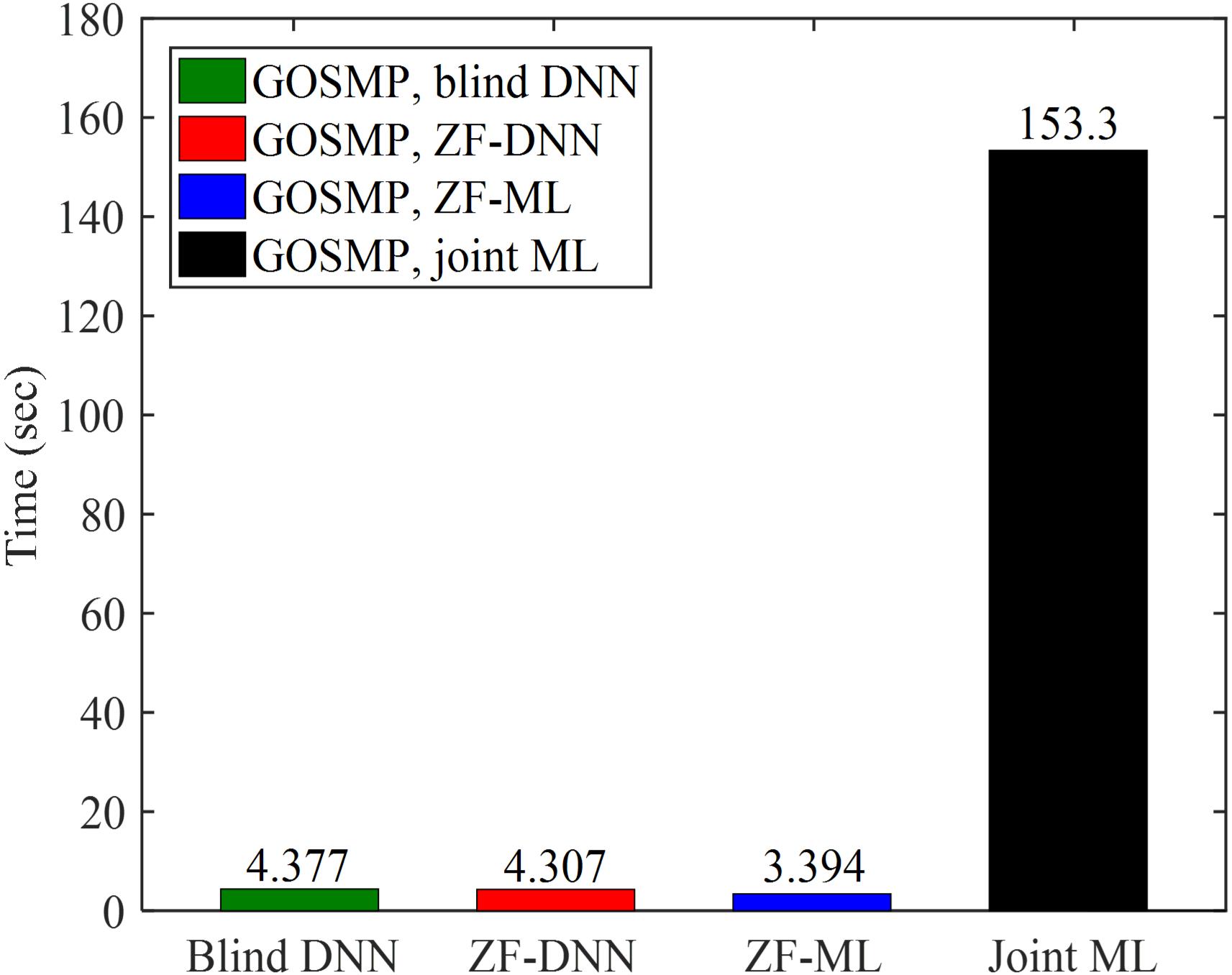}}
\caption{Computation time comparison of the proposed CSI-free blind DNN detector and three benchmark detectors for (a) GOSM and (b) GOSMP at the center of the receiving plane.}
\label{fig:Time}
\end{figure}

Due to the substantial path loss during MIMO transmission, the received signal needs to be properly amplified before it can be fed into the feed-forward DNN module. Figs. \ref{fig:BERalpha}(a) and (b) show the BER versus $\textrm{Log}_{10} \alpha$ with different transmitted SNRs for GOSM and GOSMP, respectively. For GOSM, as shown in Fig. \ref{fig:BERalpha}(a), we can observe that a feasible range of $\alpha$ is around [$10^5$, $10^7$] when the receiver is located at the center of the receiving plane. Moreover, the feasible range of $\alpha$ keeps the same for different transmitted SNR values. When the receiver is moved to the corner, the feasible range of $\alpha$ is [$10^5$, $10^8$]. For GOSMP, as shown in Fig. \ref{fig:BERalpha}(b), the same feasible range of $\alpha$ is obtained as that of GOSM when the receiver is located at the center of the receiving plane. However, the feasible range of $\alpha$ for GOSMP is only around $10^6$ when the receiver is moved to the corner. To successfully implement the proposed CSI-free blind DNN detector, the proper $\alpha$ value with respect to each receiver location is determined in advance.

\subsection{Computational Complexity}

Finally, we evaluate the computational complexity of the proposed CSI-free blind DNN detector and compare it with other benchmark detectors, in terms of computation time \cite{albinsaid2020block}. For both the CSI-free blind DNN detector and the ZF-DNN detector, once the detector has been successfully trained, it can be used for MIMO detection for a long period of time without further retraining, unless the system parameters such as receiver location have been changed \cite{wang2020deep}. Hence, only the computational complexity of the online detection process is considered for the CSI-free blind DNN detector and the ZF-DNN detector, while the complexity of the offline training process is not taken into account. Figs. \ref{fig:Time}(a) and (b) compare the computation time of the proposed blind DNN detectors and the other three benchmark detectors for GOSM and GOSMP, respectively. As we can see, for GOSM, the CSI-free blind DNN detector, the ZF-DNN detector and the ZF-ML detector require nearly the same computation time which is less than 3 seconds. However, the joint ML detector requires totally 48.42 seconds to finish the computation, which is significantly longer than that of the other three detectors. It is the same for GOSMP that the CSI-free blind DNN detector, the ZF-DNN detector and the ZF-ML detector require comparable computation time, which is much shorter than that required by the joint ML detector. Therefore, the proposed CSI-free blind DNN detector achieves near optimal BER performance as the joint ML detector, but with substantially lower computational complexity.

\section{Conclusion} 

In this paper, we have for the first time proposed a novel DeepGOMIMO framework for GOMIMO systems, where a DNN-based detector is specially designed to realize CSI-free blind detection of the received MIMO signals. The proposed CSI-free blind DNN detector contains a pre-processing module and a feed-forward DNN module, which are used to address the adverse effects of MIMO transmission and to perform joint detection of spatial and constellation information, respectively. It is shown by our simulation results that, in a typical indoor 4 $\times$ 4 MIMO-OWC system adopting both GOSM and GOSMP with unipolar non-zero 4-PAM modulation, the CSI-free blind DNN detector achieves comparable BER performance as the optimal joint ML detector, which greatly outperforms the ZF-ML detector. Moreover, the CSI-free blind DNN detector, the ZF-DNN detector and the ZF-ML detector require nearly the same computation time to perform detection, which is significantly shorter than that required by the joint ML detector. In addition, compared with the ZF-DNN detector, the CSI-free blind DNN detector can achieve improved achievable data rate and reduced communication time delay since it does not require instantaneous channel estimation to obtain accurate CSI for ZF equalization. In conclusion, our proposed DeepGOMIMO can be a potential candidate for the implementation of practical high-speed and low-complexity OWC systems.

\bibliographystyle{IEEEtran}
\bibliography{IEEEabrv,mylib}

\begin{thebibliography}{10}
\providecommand{\url}[1]{#1}
\csname url@samestyle\endcsname
\providecommand{\newblock}{\relax}
\providecommand{\bibinfo}[2]{#2}
\providecommand{\BIBentrySTDinterwordspacing}{\spaceskip=0pt\relax}
\providecommand{\BIBentryALTinterwordstretchfactor}{4}
\providecommand{\BIBentryALTinterwordspacing}{\spaceskip=\fontdimen2\font plus
\BIBentryALTinterwordstretchfactor\fontdimen3\font minus
  \fontdimen4\font\relax}
\providecommand{\BIBforeignlanguage}[2]{{%
\expandafter\ifx\csname l@#1\endcsname\relax
\typeout{** WARNING: IEEEtran.bst: No hyphenation pattern has been}%
\typeout{** loaded for the language `#1'. Using the pattern for}%
\typeout{** the default language instead.}%
\else
\language=\csname l@#1\endcsname
\fi
#2}}
\providecommand{\BIBdecl}{\relax}
\BIBdecl

\bibitem{ghassemlooy2015emerging}
Z.~Ghassemlooy, S.~Arnon, M.~Uysal, Z.~Xu, and J.~Cheng, ``Emerging optical
  wireless communications-advances and challenges,'' \emph{IEEE J. Sel. Areas
  Commun.}, vol.~33, no.~9, pp. 1738--1749, Sep. 2015.

\bibitem{cogalan2017would}
T.~Cogalan and H.~Haas, ``{Why would 5G need optical wireless
  communications?}'' in \emph{Proc. IEEE Ann. Int. Symp. Pers., Indoor Mobile
  Radio Commun. (PIMRC)}, Oct. 2017, pp. 1--6.

\bibitem{chi2020visible}
N.~Chi, Y.~Zhou, Y.~Wei, and F.~Hu, ``{Visible light communication in 6G:
  Advances, challenges, and prospects},'' \emph{IEEE Veh. Technol. Mag.},
  vol.~15, no.~4, pp. 93--102, Dec. 2020.

\bibitem{demirkol2019powering}
I.~Demirkol, D.~Camps-Mur, J.~Paradells, M.~Combalia, W.~Popoola, and H.~Haas,
  ``{Powering the Internet of Things through light communication},'' \emph{IEEE
  Commun. Mag.}, vol.~57, no.~6, pp. 107--113, May 2019.

\bibitem{chen2021noma}
C.~Chen, S.~Fu, X.~Jian, M.~Liu, X.~Deng, and Z.~Ding, ``{NOMA for
  energy-efficient LiFi-enabled bidirectional IoT communication},'' \emph{IEEE
  Trans. Commun.}, vol.~69, no.~3, pp. 1693--1706, Mar. 2021.

\bibitem{le2009100}
H.~Le~Minh, D.~O'Brien, G.~Faulkner, L.~Zeng, K.~Lee, D.~Jung, Y.~Oh, and E.~T.
  Won, ``{100-Mb/s NRZ visible light communications using a postequalized white
  LED},'' \emph{IEEE Photon. Technol. Lett.}, vol.~21, no.~15, pp. 1063--1065,
  Aug. 2009.

\bibitem{zeng2009high}
L.~Zeng, D.~C. O'Brien, H.~Le~Minh, G.~E. Faulkner, K.~Lee, D.~Jung, Y.~Oh, and
  E.~T. Won, ``{High data rate multiple input multiple output (MIMO) optical
  wireless communications using white LED lighting},'' \emph{IEEE J. Sel. Areas
  Commun.}, vol.~27, no.~9, pp. 1654--1662, Dec. 2009.

\bibitem{fath2013performance}
T.~Fath and H.~Haas, ``{Performance comparison of MIMO techniques for optical
  wireless communications in indoor environments},'' \emph{IEEE Trans.
  Commun.}, vol.~61, no.~2, pp. 733--742, Feb. 2013.

\bibitem{chen2017coverage}
C.~Chen, W.-D. Zhong, and D.~Wu, ``{On the coverage of multiple-input
  multiple-output visible light communications [Invited]},'' \emph{J. Opt.
  Commun. Netw.}, vol.~9, no.~9, pp. D31--D41, Sep. 2017.

\bibitem{chen2020user}
C.~Chen, H.~Yang, P.~Du, W.-D. Zhong, A.~Alphones, Y.~Yang, and X.~Deng,
  ``{User-centric MIMO techniques for indoor visible light communication
  systems},'' \emph{IEEE Syst. J.}, vol.~14, no.~3, pp. 3202--3213,, Sep. 2020.

\bibitem{mesleh2011optical}
R.~Mesleh, H.~Elgala, and H.~Haas, ``Optical spatial modulation,'' \emph{J.
  Opt. Commun. Netw.}, vol.~3, no.~3, pp. 234--244, Mar. 2011.

\bibitem{alaka2015generalized}
S.~Alaka, T.~L. Narasimhan, and A.~Chockalingam, ``Generalized spatial
  modulation in indoor wireless visible light communication,'' in \emph{Proc.
  IEEE Global Commun. Conf. (GLOBECOM)}, Dec. 2015, pp. 1--7.

\bibitem{wang2020constellation}
F.~Wang, F.~Yang, and J.~Song, ``{Constellation optimization under the ergodic
  VLC channel based on generalized spatial modulation},'' \emph{Opt. Exp.},
  vol.~28, no.~14, pp. 21\,202--21\,209, Jul. 2020.

\bibitem{wang2020indoor}
K.~Wang, ``{Indoor optical wireless communication system with filters-enhanced
  generalized spatial modulation and carrierless amplitude and phase (CAP)
  modulation},'' \emph{Opt. Lett}, vol.~45, no.~18, pp. 4980--4983, Sep. 2020.

\bibitem{chen2021ofdm}
C.~Chen, X.~Zhong, S.~Fu, X.~Jian, M.~Liu, H.~Yang, A.~Alphones, and H.~Y. Fu,
  ``{OFDM-based generalized optical MIMO},'' \emph{J. Lightw. Technol.},
  vol.~39, no.~19, pp. 6063--6075, Oct. 2021.

\bibitem{ozbilgin2015optical}
T.~{\"O}zbilgin and M.~Koca, ``Optical spatial modulation over atmospheric
  turbulence channels,'' \emph{J. Lightw. Technol.}, vol.~33, no.~11, pp.
  2313--2323, Jun. 2015.

\bibitem{lecun2015deep}
Y.~LeCun, Y.~Bengio, and G.~Hinton, ``Deep learning,'' \emph{Nature}, vol. 521,
  no. 7553, pp. 436--444, May 2015.

\bibitem{wang2017deep}
T.~Wang, C.-K. Wen, H.~Wang, F.~Gao, T.~Jiang, and S.~Jin, ``Deep learning for
  wireless physical layer: Opportunities and challenges,'' \emph{Chin.
  Commun.}, vol.~14, no.~11, pp. 92--111, Nov. 2017.

\bibitem{lee2018binary}
H.~Lee, I.~Lee, T.~Q. Quek, and S.~H. Lee, ``{Binary signaling design for
  visible light communication: A deep learning framework},'' \emph{Opt. Exp.},
  vol.~26, no.~14, pp. 18\,131--18\,142, Jul. 2018.

\bibitem{lu2019memory}
X.~Lu, C.~Lu, W.~Yu, L.~Qiao, S.~Liang, A.~P.~T. Lau, and N.~Chi,
  ``{Memory-controlled deep LSTM neural network post-equalizer used in
  high-speed PAM VLC system},'' \emph{Opt. Exp.}, vol.~27, no.~5, pp.
  7822--7833, Mar. 2019.

\bibitem{yang2019learning}
H.~Yang, A.~Alphones, W.-D. Zhong, C.~Chen, and X.~Xie, ``{Learning-based
  energy-efficient resource management by heterogeneous RF/VLC for
  ultra-reliable low-latency industrial IoT networks},'' \emph{IEEE Trans. Ind.
  Informat.}, vol.~16, no.~8, pp. 5565--5576, Aug. 2019.

\bibitem{wang2020deep}
T.~Wang, F.~Yang, and J.~Song, ``Deep learning-based detection scheme for
  visible light communication with generalized spatial modulation,'' \emph{Opt.
  Exp.}, vol.~28, no.~20, pp. 28\,906--28\,915, Sep. 2020.

\bibitem{wang2014demo}
Y.~Wang and N.~Chi, ``{Demonstration of high-speed 2 $\times$ 2 non-imaging
  MIMO Nyquist single carrier visible light communication with frequency domain
  equalization},'' \emph{J. Lightw. Technol.}, vol.~32, no.~11, pp. 2087--2093,
  Jun. 2014.

\bibitem{komine2004fundamental}
T.~Komine and M.~Nakagawa, ``{Fundamental analysis for visible-light
  communication system using LED lights},'' \emph{IEEE Trans. Consum.
  Electron.}, vol.~50, no.~1, pp. 100--107, Feb. 2004.

\bibitem{tavakkolnia2018ofdm}
I.~Tavakkolnia, A.~Yesilkaya, and H.~Haas, ``\protect{OFDM-based spatial
  modulation for optical wireless communications},'' in \emph{Proc. IEEE
  Globecom Workshops (GC Wkshps)}, Dec. 2018, pp. 1--6.

\bibitem{chen2021deep}
C.~Chen, L.~Zeng, X.~Zhong, S.~Fu, M.~Liu, and P.~Du, ``{Deep learning-aided
  OFDM-based generalized optical quadrature spatial modulation},'' 2021.
  [Online]. Available: $\textrm{https://arxiv.org/abs/2106.12770}$.

\bibitem{ying2015joint}
K.~Ying, H.~Qian, R.~J. Baxley, and S.~Yao, ``{Joint optimization of precoder
  and equalizer in MIMO VLC systems},'' \emph{IEEE J. Sel. Areas Commun.},
  vol.~33, no.~9, pp. 1949--1958, Sep. 2015.

\bibitem{albinsaid2020block}
H.~Albinsaid, K.~Singh, S.~Biswas, C.-P. Li, and M.-S. Alouini, ``Block deep
  neural network-based signal detector for generalized spatial modulation,''
  \emph{IEEE Commun. Lett.}, vol.~24, no.~12, pp. 2775--2779, Dec. 2020.

\end{thebibliography}

\end{document}